\documentclass[runningheads,fleqn]{svmult}
\usepackage{makeidx}   % allows index generation
\usepackage{graphicx}  % standard LaTeX graphics tool
\usepackage{subeqnar}  % subnumbers individual equations
\usepackage{multicol}  % used for the two-column index
\usepackage{taphys}    % flushleft layout of captions etc.
\makeindex             % used for the subject index
\usepackage[]{amsbsy}
\newcommand{\ff}[1]{{\boldsymbol #1}}
\begin{document}
\title*{Dynamical Variational Principles for Strongly Correlated Electron Systems}
\toctitle{Dynamical Variational Principles for Strongly Correlated Electron Systems}
% allows explicit linebreak for the table of content
%
%
\titlerunning{Dynamical Variational Principles}
% allows abbreviation of title, if the full title is too long
% to fit in the running head
%
\author{Michael Potthoff}
\authorrunning{Michael Potthoff}
% if there are more than two authors,
% please abbreviate author list for running head
%
%
\institute{Institut f\"ur Theoretische Physik und Astrophysik, Universit\"at 
W\"urzburg, Am Hubland, 97074 W\"urzburg, Germany}

\maketitle              % typesets the title of the contribution

%%%%%%%%%%%%%%%%%%%%%%%%%%%%%%%%%%%%%%%%%%%%%%%%%%%%%%%%%%%%%%%%%%%%%%%%%%%%%%%%%%%%%%%%%%%%%%%%%%%
\begin{abstract}
The self-energy-functional approach (SFA) is discussed in the context of 
different variational principles for strongly correlated electron systems.
Formal analogies between static and dynamical variational approaches, different
types of approximation strategies and the relations to density-functional and 
dynamical mean-field theory are emphasized.
The discussion elucidates the strengths of the SFA in the construction of new 
non-perturbative approximations but also the limitations of the approach and
thereby opens up future perspectives.
\end{abstract}

Variational approaches have a long and successful tradition in the theory of 
condensed-matter systems as they offer a smart, controlled and systematic
way to treat the problem of electron-electron interaction.
A well-known variational approach is Hartree-Fock (HF) theory.
It is based on the Rayleigh-Ritz principle and provides a practicable and 
consistent mean-field description of an interacting electron system.
As quantum fluctuations are neglected completely, HF theory must be classified
as a {\em static} mean-field approximation.
This may be contrasted with dynamical mean-field theory (DMFT) \cite{GKKR96,MV89} 
which includes temporal fluctuations in the mean-field picture.
The DMFT, however, cannot be derived from the Ritz principle.
It must be constructed from some {\em dynamical} variational principle which involves  
a dynamical (i.e.\ time- or frequency-dependent) quantity as the basic object.
Dynamical variational principles have already been suggested in 
the sixties \cite{LW60,BK61} but, compared to the Ritz principle, were employed 
with rather limited success only.
This brings up the following questions:
What are the similarities and the differences between different variational 
principles and approximation strategies?
How can the DMFT be considered as an approximation within a variational concept?
Can dynamical variational principles be used for constructing practicable
and non-perturbative approximations different from the DMFT?
An attempt to answer these questions straightforwardly leads to the
self-energy-functional approach (SFA) \cite{Pot03a} suggested recently.
The purpose of this paper is to discuss different variational approaches
and to place the SFA into this context with the objective to explore possible 
future developments.

%*******************************************************************************
\section{Variational Principles and Approximation Strategies}
%*******************************************************************************

Consider a many-electron system in the volume $V$, at temperature $T$ and 
with chemical potential $\mu$. 
It is characterized by a Hamiltonian $H_{\ff t, \ff U} = H_0(\ff t) + H_1(\ff U)$ 
consisting 
of a one-particle and an interaction term $H_0$ and $H_1$, respectively, which 
depend on the ``model parameters'' $\ff t$ and $\ff U$ (a matrix notation is used).
In second-quantized form,
\begin{equation}
  H_{\ff t, \ff U} = \sum_{\alpha\beta} t_{\alpha\beta} \:
  c_{\alpha}^\dagger c_{\beta}
  + \frac{1}{2} 
  \sum_{\alpha\beta\gamma\delta} U_{\alpha\beta\delta\gamma} \:
  c_{\alpha}^\dagger c^\dagger_{\beta} c_{\gamma} c_{\delta} \: ,
\label{eq:ham}
\end{equation}
where an index (e.g.\ $\alpha$) refers to the states of a one-particle basis.

The characteristic of a variational approach is a certain physical quantity 
$\ff X$ to be varied, as e.g.\ the statistical operator, the electron density, 
the (local) Green's function, the self-energy etc. Clearly, at equilibrium
this quantity will depend on the model parameters: $\ff X_{\ff t, \ff U}$
(and on $V, T$ and $\mu$).

In a variational approach, the quantity is considered to be a variable.
The first task is to express a thermodynamical potential $\Omega$ (the 
grand potential, for example) as a function(al) of this variable:
$\Omega_{\ff t, \ff U}[\ff X]$.
As this functional is characteristic for the model system (\ref{eq:ham}), it
will depend on the model parameters.
At the equilibrium or ``physical'' value, i.e.\ at $\ff X=\ff X_{\ff t, \ff U}$, one 
must have $\Omega_{\ff t, \ff U}[\ff X_{\ff t, \ff U}] = \Omega_{\ff t, \ff U}$ 
where $\Omega_{\ff t, \ff U} = - T \ln \mbox{tr} \exp (- (H_0(\ff t) + H_1(\ff U) - \mu N)/T)$.

Furthermore, the functional $\Omega_{\ff t, \ff U}[\ff X]$ should be constructed 
such that it becomes stationary at the physical value: 
$\delta \Omega_{\ff t, \ff U}[\ff X=\ff X_{\ff t, \ff U}]= 0$.
This {\em variational principle} determines $\ff X_{\ff t, \ff U}$ once the functional 
is known.
Note that the domain of the functional must be specified in addition since in most cases 
$\ff X_{\ff t, \ff U}$ must satisfy some constraint or sum rule or normalization condition.

Even if the functional is known, however, it is usually 
impossible to evaluate $\Omega_{\ff t, \ff U}[\ff X]$ for a given $\ff X$, and one
has to resort to approximations. 
One may distinguish between three types of approximation strategies:

In a {\em type-I approximation} one derives the Euler equation
$\delta \Omega_{\ff t, \ff U}[\ff X] / \delta \ff X = 0$ first and then
chooses (a physically motivated) simplification of the equation
afterwards to render the determination of $\ff X_{\ff t, \ff U}$ possible.
This is the most general but worst type, as normally the approximated
Euler equation no longer derives from some approximate functional.
This may result in thermodynamical inconsistencies.

A {\em type-II approximation} modifies the form of the functional dependence,
$\Omega_{\ff t, \ff U}[\cdots] \to \widetilde{\Omega}_{\ff t, \ff U}[\cdots]$,
to get a simpler one that allows for a solution of the resulting Euler equation
$\delta \widetilde{\Omega}_{\ff t, \ff U}[\ff X] / \delta \ff X = 0$.
This type is more particular but yields a thermodynamical potential consistent 
with $\ff X_{\ff t, \ff U}$.
Generally, however, it is not easy to find a sensible approximation of a 
functional form.

Finally, in a {\em type-III approximation} one restricts the domain of the 
functional. 
The restriction comes in addition to those conditions that are physically 
necessary anyway (e.g.\ normalizations) and requires a precise definition of the 
domain.
This type is most specific and, from a conceptual point of view, should be 
preferred as compared to type-I or type-II 
approximations as the exact functional form is retained.
In addition to conceptual clarity and thermodynamical consistency, type-III approximations
are truely systematic since 
improvements can be obtained by an according extension of the domain.
Note that any type-III approximation can always be understood as a type-II one 
(and type-II approximations as type-I) but not vice versa.

%*******************************************************************************
\section{Various Variational Approaches}
%*******************************************************************************

In the following different variational principles and possible
approximations shall be discussed according to this scheme, starting with
Hartree-Fock and density-functional theory as illuminating examples.

%*******************************************************************************
\subsection{Ritz Variational Approach}
%*******************************************************************************

In the Ritz variational approach the ground-state energy is considered as a 
functional of the quantum state $|\Psi \rangle$.
There is a generalization of the Ritz principle to arbitrary temperatures by
Mermin \cite{Mer65}.
Here the basic variable is the statistical operator $\rho$ characterizing the
system's (mixed) state, and the grand potential as a functional of $\rho$ reads:
\begin{equation}
  \Omega_{\ff t, \ff U}[\rho] = {\rm tr} (\rho (H_{\ff t, \ff U} - \mu N + T \ln \rho)) \: .
\label{eq:ritz}  
\end{equation}
Following the classical calculation of Gibbs, it can easily be shown \cite{Mer65} 
that on the set of normalized and positive definite operators, stationarity
of the functional (\ref{eq:ritz}) is achieved for $\rho = \rho_{\ff t, \ff U} 
= e^{-(H_{\ff t, \ff U}-\mu N)/T} / {\rm tr} ( e^{-(H_{\ff t, \ff U}-\mu N)/T} )$.
One also has $\Omega_{\ff t, \ff U}[\rho_{\ff t, \ff U}] = \Omega_{\ff t, \ff U}$.
An additional feature of the functional (\ref{eq:ritz}) consists in the fact that 
$\Omega_{\ff t, \ff U} [\rho] \ge \Omega_{\ff t, \ff U} [\rho_{\ff t, \ff U}]$
for any $\rho$.
This ``upper-bound property'' is extremely helpful but specific to the Ritz principle.
 
For a many-electron system and an arbitrary $\rho$, the computation of the trace 
in Eq.\ (\ref{eq:ritz}) is an exponentially hard problem.
A nice type-III approximation is the HF approach: 
Here the variational search is restricted to the subclass of {\em disentangled statistical 
operators}, i.e.\ statistical operators corresponding to independent-electron states.
This can be made precise by introducing the important concept of a reference system:

A reference system is a system with a different (microscopic) Hamiltonian 
$H_{\ff t', \ff U'}$ ($\ff t' \ne \ff t$, $\ff U' \ne \ff U$) 
but with a macroscopic state characterized by the same values of the thermodynamic state 
variables as the original system (\ref{eq:ham}): $V'=V$, $T'=T$ and $\mu'=\mu$. 
The sole purpose of the reference system is to specify the domain of the functional
(\ref{eq:ritz}):
Trial statistical operators are taken from the reference system, $\rho = \rho_{\ff t', \ff U'}$,
and are varied by varying the parameters $\ff t'$ and $\ff U'$ within a certain
subspace.
Hence, the choice of the reference system (and the parameter subspace)
defines the approximation.

The HF approximation is given by the choice $\ff U'=0$ and $\ff t'$ arbitrary,
i.e.\ by trial states $\rho_{\ff t', 0} =  e^{-(H_{\ff t', 0}-\mu N)/T} / Z_{\ff t', 0}$.
Inserting into Eq.\ (\ref{eq:ritz}) yields
\begin{equation}
  \Omega_{\ff t, \ff U}[\rho_{\ff t',0}]
   = 
   \Omega_{\ff t', 0}
   + {\rm tr} (\rho_{\ff t',0} (H_0(\ff t) + H_1(\ff U) - H_0(\ff t')) \: .
\end{equation}
The remaining trace can be computed easily using Wick's theorem as $\rho_{\ff t',0}$
derives from a non-interacting Hamiltonian.
The variational parameters $\ff t'$ are fixed by the conditions 
$\delta \Omega_{\ff t, \ff U}[\rho_{\ff t',0}] / \delta \ff t' =0$.
These are exactly the well-known HF equations as can be seen by some straightforward 
manipulations.

One learns that type-III approximations can be constructed conveniently by the
{\em concept of a reference system}. 
On the one hand, the reference system should comprise a large space of parameters 
$\ff t'$ and $\ff U'$ to generate a powerful approximation.
On the other hand, the parameter space must be restricted strongly to keep the 
calculations feasible.

%*******************************************************************************
\subsection{Density-Functional Approach}
%*******************************************************************************

For a many-electron system the statistical operator or, at $T=0$, the ground-state
wave function actually is an object that is by far too complex.
The {\em relevant} information is much more efficiently stored in 
integral quantities, such as the electron density.
This is the starting point of density-functional theory (DFT) \cite{HK64,KS65,DG90}.
The density, i.e.\ the quantum-statistical {\em average} of the one-particle density 
operator $n(\ff r) = \mbox{tr} (\rho \hat{n}(\ff r))$,
represents the basic variable.
Normally DFT aims at the inhomogeneous electron gas at $T=0$ but can also be applied 
to discrete lattice models \cite{SGN95} and finite temperatures \cite{Mer65}.

The grand potential $\Omega_{\ff t, \ff U}$ obviously depends on the model 
parameters.
Due to the Hohenberg-Kohn theorem \cite{HK64}, however, it can also be considered
as a functional of the density $\ff n$ which is stationary at the physical density:
$\delta \Omega_{\ff t, \ff U}[\ff n]=0$ for $\ff n = \ff n_{\ff t, \ff U}$.
Furthermore, if evaluated at $\ff n = \ff n_{\ff t, \ff U}$, it yields the exact 
grand potential: $\Omega_{\ff t, \ff U}[\ff n_{\ff t, \ff U}]=\Omega_{\ff t, \ff U}$.
Keeping the notations introduced above, $\ff n$ is a matrix with 
$n_{\alpha\beta}=\mbox{tr} (\rho \, c_\alpha^\dagger c_\beta)$, and the functional
reads (cf.\ Refs.\ \cite{Mer65,SGN95}):
\begin{equation}
  \Omega_{\ff t, \ff U}[\ff n] = \mbox{tr}(\ff t \, \ff n) + F_{\ff U}[\ff n] \: .
\label{eq:dft}
\end{equation}
Here the trace refers to the one-particle orbitals $\alpha$, and $F_{\ff U}[\ff n]$
is a {\em universal} functional, i.e.\ it depends on the interaction parameters only.
Using the Kohn-Sham idea \cite{KS65,SGN95}, the resulting Euler equation has the 
form of a one-particle Schr\"odinger equation.

The variational principle $\delta \Omega_{\ff t, \ff U}[\ff n]=0$ is rigorous but 
cannot be evaluated as $F_{\ff U}[{\ff n}]$ is generally unknown (after separating 
the Hartree and a kinetic-energy term, the remaining exchange-correlation functional 
is not explicit).
Due to the universality of $F_{\ff U}[{\ff n}]$, however, the density-functional 
for a reference system with modified one-particle parameters $\ff t'$ reads
$\Omega_{\ff t', \ff U}[\ff n] = \mbox{tr}(\ff t' \, \ff n) + F_{\ff U}[\ff n]$,
and thus 
$\Omega_{\ff t, \ff U}[\ff n_{\ff t', \ff U}] = \Omega_{\ff t', \ff U}
+ \mbox{tr}((\ff t - \ff t') \ff n_{\ff t', \ff U})$
which can be exploited for a type-III approximation.
Choosing as a reference system $H_{\ff t', \ff U}$ the homogeneous electron gas,
however, turns out to be too restrictive, as this implies a spatially constant 
density.
The {\em local} density approximation (LDA) \cite{HK64,KS65}, on the other hand, 
has proven to be very successful \cite{DG90}.
At least for systems with weakly varying density it is well justified.
The LDA, however, is no longer a type-III approximation but a type-II one as the
{\em form} of the (exchange-correlation part of the) functional $F_{\ff U}[\ff n]$ 
is approximated to have a local dependence on the density only.

As the proof of the Hohenberg-Kohn theorem is based on the Ritz principle \cite{HK64},
the upper-bound property is transferred to the exact functional (\ref{eq:dft}), i.e.\
$\Omega_{\ff t, \ff U}[\ff n] \ge \Omega_{\ff t, \ff U}$ for any $\ff n$, but 
is lost within the LDA due to the type-II character of the approximation.

%*******************************************************************************
\subsection{Time-Dependent DFT}
%*******************************************************************************

The weak point of the DFT consists in its inability to describe excitations
(see, however, Ref.\ \cite{SK66}). 
This is due to the fact that the Hohenberg-Kohn variational principle is built
on the static electron density.
Information on excitation properties is contained in dynamic response functions 
which are accessible in principle via time-dependent density-functional theory 
(TD-DFT) \cite{RG84}. 
In TD-DFT one considers a situation with a time-dependent Hamiltonian 
and focuses on the time-dependent density 
$n(\ff r,t) = \langle \Psi(t) | \hat{n}(\ff r) | \Psi(t) \rangle$ resulting
from a solution $| \Psi(t) \rangle$ of the time-dependent Schr\"odinger equation
as the basic variable.
Here the action $A = \int dt \langle \Psi(t) | i \partial/\partial t - H(t) 
| \Psi(t) \rangle$ can be understood as a functional of $n(\ff r,t)$,
\begin{equation}
A_{\ff t, \ff U}[{\ff n}] =  - \int_{t_0}^{t_1} dt \: \mbox{tr} ( \ff n(t) \ff t(t) )
+
B_{\ff U}[{\ff n}] \: ,
\end{equation}
where $\ff t(t)$ are explicitly time-dependent one-particle parameters.
Contrary to usual DFT, the variational principle $\delta A_{\ff t, \ff U} [\ff n] = 0$ 
does not derive from the Ritz principle, and consequently there is no upper-bound 
property available.
Type-II approximations can be constructed by approximating the universal 
but unknown part $B_{\ff U}[{\ff n}]$ of the functional to make it explicit.
Far from equilibrium, however, there is no general recipe.

%*******************************************************************************
\subsection{Dynamical Variational Principle}
%*******************************************************************************

In the linear-response regime close to equilibrium, excitations are 
described by Green's functions.
The one-electron Green's function 
$G_{\alpha\beta}(\omega) = \langle \langle c_\alpha ; c_\beta^\dagger \rangle \rangle_\omega$
is the basic quantity in the dynamical variational approach of Luttinger, Ward, Baym and 
Kadanoff \cite{LW60,BK61}.
Employing a coupling-constant integration \cite{LW60}, the grand potential can be 
understood as a functional of $\ff G$:
\begin{equation}
\Omega_{\ff t, \ff U}[\ff G] = \mbox{Tr} \ln \ff G - \mbox{Tr} (( \ff G_{\ff t,0}^{-1} - \ff G^{-1} ) \ff G)
+ \Phi_{\ff U}[\ff G] \; ,
\label{eq:bk}
\end{equation}
where $\mbox{Tr} = T \sum_n e^{i\omega_n 0^+} \mbox{tr}$ and $\omega_n = (2n+1)\pi T$ are
fermion Matsubara frequencies.
Furthermore, $\ff G_{\ff t,0} = (\omega + \mu - \ff t)^{-1}$ is the $\ff U=0$ (free) 
Green's function and $\Phi_{\ff U}[\ff G]$ the (universal) Luttinger-Ward (LW) functional 
defined as the sum of all dressed closed skeleton diagrams \cite{LW60}.
By construction, $\Omega_{\ff t, \ff U}[\ff G_{\ff t, \ff U}] = \Omega_{\ff t, \ff U}$.
In arbitrary order in perturbation theory one has
$\delta \Phi_{\ff U}[\ff G] / \delta \ff G = T \ff \Sigma_{\ff U}[\ff G]$.
Therewith, the Euler equation $\delta \Omega_{\ff t, \ff U}[\ff G] / \delta \ff G = 0$ is 
given by $\ff G^{-1} - \ff G_{\ff t,0}^{-1} + \ff \Sigma_{\ff U}[\ff G] = 0$ which is 
Dyson's equation.
This shows that $\Omega_{\ff t, \ff U}[\ff G]$ is stationary at the physical Green's function
$\ff G = \ff G_{\ff t, \ff U}$. 

The LW functional is formally given by a diagrammatic sum that cannot be carried out in practice.
A self-evident type-II strategy is to sum up a suitable subclass of diagrams to obtain an approximate 
but explicit expression for $\Phi_{\ff U}[\ff G]$.
In this way the HF approximation can be recovered but there are also new approximations like
the fluctuation-exchange approximation \cite{BSW89,BS89}.
These ``conserving approximations'', however, are necessarily restricted to the weak-coupling
regime.

A type-III approximation, on the other hand, would be non-perturbative by construction. 
Consider a reference system with modified one-particle parameters: 
$H_{\ff t', \ff U} = H_0(\ff t') + H_1(\ff U)$. This defines the domain of the functional
(\ref{eq:bk}) to consist of Green's functions $\ff G_{\ff t', \ff U}$ with arbitrary $\ff t'$.
The interaction is kept fixed ($\ff U' = \ff U$).
To evaluate the functional (\ref{eq:bk}) at $\ff G_{\ff t', \ff U}$ requires the evaluation
of $\Phi_{\ff U}[\ff G_{\ff t',\ff U}]$, in particular.
Due to the universality of $\Phi_{\ff U}[\cdots]$ (no $\ff t$ dependence) and due to the choice
$\ff U' = \ff U$, one has 
$\Phi_{\ff U}[\ff G_{\ff t',\ff U}] = \Omega_{\ff t', \ff U} - \mbox{Tr} \ln \ff G_{\ff t',\ff U} 
+ \mbox{Tr} (( \ff G_{\ff t',0}^{-1} - \ff G_{\ff t',\ff U}^{-1} ) \ff G_{\ff t',\ff U})$.
Thus,
\begin{equation}
\Omega_{\ff t, \ff U}[\ff G_{\ff t',\ff U}]
=
\Omega_{\ff t', \ff U}
-
\mbox{Tr} ( \ff G_{\ff t,0}^{-1} \ff G_{\ff t',\ff U})
+
\mbox{Tr} ( \ff G_{\ff t',0}^{-1} \ff G_{\ff t',\ff U})
\: .
\label{eq:dyn}
\end{equation}
Hence, on any domain specified by a suitable subspace of one-particle parameters 
$\ff t'$ which renders the solution of the reference system possible (for fixed $\ff U$),
the functional (\ref{eq:bk}) can be evaluated exactly.
A possible (but oversimplified) example is the choice $\ff t' = 0$. 
It reduces the reference model to the atomic limit where the computation 
of $\ff G_{\ff t',\ff U}$ and $\Omega_{\ff t', \ff U}$ in (\ref{eq:dyn}) is easy.
Cluster approximations represent straightforward generalizations of this example. 

Unfortunately, this type-III approach for Eq.\ (\ref{eq:bk}) yields nothing new:
Since
$\mbox{Tr} ( \ff G_{\ff t,0}^{-1} - \ff G_{\ff t',0}^{-1} ) \ff G_{\ff t',\ff U} 
=
\mbox{tr} ( \ff t - \ff t') \ff n_{\ff t',\ff U} 
$
with the one-electron density of the reference system 
$\ff n_{\ff t',\ff U} = T \sum_n e^{i\omega_n 0^+}\ff G_{\ff t',\ff U}(i\omega_n)$,
one gets
$\Omega_{\ff t, \ff U}[\ff G_{\ff t',\ff U}]=\Omega_{\ff t, \ff U}[\rho_{\ff t',\ff U}]$
with $\Omega_{\ff t, \ff U}[\rho]$ given by Eq.\ (\ref{eq:ritz}),
i.e.\ the same as in the Ritz variational approach.
Interestingly, this implies that upper bounds for the grand potential can be 
obtained, i.e.\
$\Omega_{\ff t, \ff U}[\ff G_{\ff t',\ff U}] \ge \Omega_{\ff t, \ff U}$ 
for arbitrary $\ff t'$.

%*******************************************************************************
\subsection{Dynamical Mean-Field Approach}
%*******************************************************************************

Equipped with these insights, one can address the question of {\em deriving the DMFT
from a variational principle}. Originally, the DMFT was introduced as the exact 
theory of lattice models with local (Hubbard-type) interactions in infinite 
spatial dimensions $D=\infty$ \cite{MV89}.
Later on, it was recognized \cite{GK92a,Jar92} that in $D=\infty$ the lattice 
model $H_{\ff t,\ff U}$ can be self-consistently mapped onto an impurity model 
$H_{\ff t',\ff U}$ with the same interaction $\ff U$.
Using this self-consistent mapping procedure as an approximation
(``dynamical mean-field approximation''), one can treat lattice models 
for any finite $D$.

Instead of considering Dyson's equation in the form,
$\ff G = (\ff G_{\ff t,0}^{-1} - \ff \Sigma_{\ff U}[\ff G])^{-1}$
(with $\ff \Sigma_{\ff U}[\ff G]) = (1/T) \, \delta \Phi_{\ff U}[\ff G]) / \delta \ff G$),
which is solved by the exact $\ff G_{\ff t, \ff U}$, the
DMFT considers the following simplified equation between {\em local} quantities
at lattice site $i$:
\begin{equation}
(\ff G)_{ii} = (\ff G_{\ff t,0}^{-1} - \widetilde{\ff \Sigma}_{\ff U}[\ff G])_{ii}^{-1}
\: .
\label{eq:dmftsc}
\end{equation}
Here $\widetilde{\ff \Sigma}_{\ff U}[\ff G]$ is the derivative of the LW functional
but with {\em local} vertices only as it is the case for an impurity model.
Clearly, this is a type-I approximation.
Eq.\ (\ref{eq:dmftsc}) is often called the DMFT self-consistency condition.
This is because its solution is achieved by an iterative procedure in practice:
Starting with a guess for $\ff \Sigma$, one computes the local lattice Green's function
as $(\ff G)_{ii} = (\ff G_{\ff t,0}^{-1} - \ff \Sigma)_{ii}^{-1}$ at first. 
This is not yet a solution of Eq.\ (\ref{eq:dmftsc}) since in general 
$\ff \Sigma \ne \widetilde{\ff \Sigma}_U[\ff G]$ for this $\ff G$.
For the necessary update of $\ff \Sigma$ define 
$(\ff G_{\ff t',0})_{ii} = 1 / (1/(\ff G)_{ii} + (\ff \Sigma)_{ii})$.
Assuming that $\ff G_{\ff t',0}$ can be understood as the free impurity Green's 
function of an impurity model $H_{\ff t',\ff U}$ for some $\ff t'$, the 
(numerical) solution of the impurity problem yields a new 
$\ff \Sigma = \widetilde{\ff \Sigma}_U[(\ff G)_{ii}] = \widetilde{\ff \Sigma}_U[\ff G]$.
Iteration of this cycle until self-consistency yields a solution $\ff G$ of 
Eq.\ (\ref{eq:dmftsc}). Note that the resulting DMFT self-energy is local.

Here, the question is whether Eq.\ (\ref{eq:dmftsc}) can be interpreted as an Euler equation
of some variational principle.
Starting with the functional (\ref{eq:bk}), one can try a type-II approximation by 
replacing $\Phi_{\ff U}[\ff G]$ with the LW functional of the impurity model
$\widetilde{\Phi}_{\ff U}[\ff G]$. 
This implies $(1/T) \delta \widetilde{\Phi}_{\ff U}[\ff G] /
\delta \ff G = \widetilde{\ff \Sigma}_U[\ff G]$, and the resulting Euler equation
reads: $\ff G^{-1} = \ff G_{\ff t,0}^{-1} - \widetilde{\ff \Sigma}_{\ff U}[\ff G]$.
This equation is easily seen to be equivalent with Eq.\ (\ref{eq:dmftsc}) since
$\widetilde{\ff \Sigma}_U[(\ff G)_{ii}] = \widetilde{\ff \Sigma}_U[\ff G]$ by 
definition.
Hence, DMFT can be understood as a type-II approximation.

Another functional has been suggested recently \cite{CK01}:
\begin{equation}
\Omega_{\ff t, \ff U}[\ff G] = \mbox{Tr} \ln \frac{1}{\ff G_{\ff t,0}^{-1} - \ff \Sigma_{\ff U}[\ff G]} 
- \mbox{Tr} ( \ff \Sigma_{\ff U}[\ff G] \ff G ) 
+ \Phi_{\ff U}[\ff G]
\: .
\label{eq:ck}
\end{equation}
Clearly, $\Omega_{\ff t, \ff U}[\ff G_{\ff t, \ff U}] = \Omega_{\ff t, \ff U}$,
and furthermore the corresponding Euler equation,
$
\left( (\ff G_{\ff t,0}^{-1} - \ff \Sigma_{\ff U}[\ff G])^{-1} - \ff G \right) \cdot 
(\delta \ff \Sigma_{\ff U}[\ff G] / \delta \ff G) = 0
$,
is equivalent with Dyson's equation,
$\ff G = (\ff G_{\ff t,0}^{-1} - \ff \Sigma_{\ff U}[\ff G])^{-1}$
(assuming local invertibility of the functional $\ff \Sigma_{\ff U}[\ff G]$).
The functional (\ref{eq:ck}) therefore yields a valid variational principle. 
As a type-II approximation, one may replace $\Phi_{\ff U}[\ff G] \to \widetilde{\Phi}_{\ff U}[\ff G]$
and $\ff \Sigma_{\ff U}[\ff G] \to \widetilde{\ff \Sigma}_{\ff U}[\ff G] = (1/T) 
\delta \widetilde{\Phi}_{\ff U}[\ff G] /
\delta \ff G$ 
in the functional (\ref{eq:ck}).
The resulting Euler equation is equivalent with the DMFT self-consistency 
equation (\ref{eq:dmftsc}) which implies that DMFT can also be understood 
as a type-II approximation to the functional (\ref{eq:ck}).

Attempts to prove that a stationary point of the type-II approximated functionals 
(\ref{eq:bk}) or (\ref{eq:ck}) must be an extremum have failed \cite{CK01}.
Furthermore, while (as shown above) a type-III approximation to the principle based 
on Eq.\ (\ref{eq:bk}) with $\ff U$ fixed yields upper bounds to the grand potential, 
the DMFT cannot be 
obtained as a type-III approximation starting from Eqs.\ (\ref{eq:bk}) or (\ref{eq:ck}):
Choosing an impurity model $H_{\ff t', \ff U}$ as a reference system to generate trial 
Green's functions and to define a restricted domain of the functional (\ref{eq:bk}) or 
(\ref{eq:ck}), respectively, concurrently means that the optimal Green's function will 
be local.
This is obviously a very poor approximation for the Green's function of a lattice 
model and differs from the DMFT result.
The discussion shows that the question whether or not the DMFT grand potential is an 
upper bound to the true grand potential is still open.

%*******************************************************************************
\subsection{Functionals of the Local Green's Function}
%*******************************************************************************

It is also possible \cite{CK00,Geo04} to focus on the local Green's function 
$\ff G^{\rm (loc)} = (\ff G)_{ii}$ (instead of the full $\ff G$) and to set
up a variational principle of the form
\begin{equation}
  \delta \Omega_{\ff t, \ff U}[\ff G^{\rm (loc)}] = 0 \: .
\end{equation}
A functional which is stationary at the physical 
$\ff G^{\rm (loc)} = \ff G_{\ff t,\ff U}^{(\rm loc)}$ and which yields
$\Omega_{\ff t, \ff U}[\ff G_{\ff t,\ff U}^{(\rm loc)}] = \Omega_{\ff t, \ff U}$
can be constructed order by order in the interaction strength \cite{CK00}.
Unfortunately, the diagrammatic formalism is much more cumbersome as compared to 
the construction of the LW functional.
As is shown in Ref.\ \cite{Geo04}, the dynamical mean-field approximation is 
equivalent with a simple ($\ff U$-independent) approximation to the kinetic-energy 
part of the functional. 
So the DMFT appears as a type-II approximation again.
 
%*******************************************************************************
\subsection{Self-Energy-Functional Approach}
%*******************************************************************************

The motivation to characterize the DMFT as a type-III approximation is the 
following: 
If it is possible to recover the DMFT merely by restricting the domain of the 
functional corresponding to an exact variational principle, different choices 
of the domain will place the DMFT in a systematic series of different and possibly 
new approximations which, as the DMFT, are all non-perturbative and thermodynamically 
consistent.

For this purpose it is helpful to focus on the self-energy.
Within the self-energy-functional approach (SFA) \cite{Pot03a}, the
self-energy functional
\begin{equation}
\Omega_{\ff t, \ff U}[\ff \Sigma] = \mbox{Tr} \ln \frac{1}{\ff G_{\ff t,0}^{-1} - \ff \Sigma} 
+ F_{\ff U}[\ff \Sigma] 
\label{eq:sfa}
\end{equation}
is considered.
Here, $F_{\ff U}[\ff \Sigma] = \Phi_{\ff U}[\ff G[\ff \Sigma]] - 
\mbox{Tr} ( \ff \Sigma \, \ff G_{\ff U}[\ff \Sigma] )$ 
is the Legendre transform of the Luttinger-Ward functional
which is well defined provided that the functional $\ff \Sigma_{\ff U}[\ff G]$ 
is invertible locally.
$F_{\ff U}[\ff \Sigma]$ is universal (independent of $\ff t$) by construction
and $- (1/T) \delta F_{\ff U}[\ff \Sigma] / \delta \ff \Sigma = \ff G_{\ff U}[\ff \Sigma]$
which is the inverse of the functional $\ff \Sigma_{\ff U}[\ff G]$.
Obviously, $\Omega_{\ff t, \ff U}[\ff \Sigma_{\ff t, \ff U}] = \Omega_{\ff t, \ff U}$.
The Euler equation $\delta \Omega_{\ff t, \ff U}[\ff \Sigma] / \delta \ff \Sigma = 0$
is given by $(\ff G_{\ff t,0}^{-1} - \ff \Sigma)^{-1} = \ff G[\ff \Sigma]$ and equivalent
with Dyson's equation.

To construct a type-III approximation, a reference system $H_{\ff t', \ff U} = 
H_0(\ff t') + H_1(\ff U)$ with unchanged interaction part is considered.
The one-particle parameters $\ff t'$ are taken such that the different ``correlated''
sites (non-zero on-site interaction) are decoupled.
Instead, $\ff t'$ shall include an arbitrary hopping to ``bath'' sites (zero on-site 
interaction) with arbitrary one-particle energies.
In case of the Hubbard model on a lattice with $L$ sites, the corresponding reference 
system constructed in this way is a set of $L$ decoupled single-impurity Anderson models
(which in case of translational symmetry are equivalent).
Trial self-energies $\ff \Sigma_{\ff t', \ff U}$ are local by construction.
The Euler equation resulting from this type-III approach reads
$\partial \Omega_{\ff t,\ff U}[\ff \Sigma_{\ff t', \ff U}] / \partial \ff t'=0$, i.e.:
\begin{equation}
%   T \sum_{\omega_n}
   \left(
   (\ff G_{\ff t,0}^{-1} - \ff \Sigma_{\ff t', \ff U})^{-1}
   - \ff G_{\ff U}[\ff \Sigma_{\ff t', \ff U}]
   \right) \cdot
   \frac{\partial \ff \Sigma_{\ff t', \ff U}}{\partial {{\bf t}'}}
   = 0  \: . 
\label{eq:euler}
\end{equation}
Now let $\ff t'$ (the bath parameters) be such that $\ff G_{\ff t',\ff U}$ solves the DMFT 
self-consistency condition (\ref{eq:dmftsc}).
Since $\widetilde{\ff \Sigma}_{\ff U}[\ff G_{\ff t',\ff U}] = \ff \Sigma_{\ff t', \ff U}$,
one has
$(\ff G_{\ff t',\ff U})_{ii} = (\ff G_{\ff t,0}^{-1} - \ff \Sigma_{\ff t', \ff U})^{-1}_{ii}$.
Hence, this $\ff t'$ solves Eq.\ (\ref{eq:euler}).
(Note that $\partial \ff \Sigma_{\ff t', \ff U} / \partial {{\bf t}'}$ is local). 
So {\em by a restriction of the domain of the self-energy functional} (\ref{eq:sfa}) to local 
self-energies, 
{\em the DMFT is characterized as a type-III approximation}.

Interestingly, a type-II approximation does not work:
A replacement of the form $F_{\ff U}[\ff \Sigma] \to \widetilde{F}_{\ff U}[\ff \Sigma]$
in Eq.\ (\ref{eq:sfa})
yields the Euler equation
$\widetilde{\ff G}_{\ff U}[\ff \Sigma] = (\ff G_{\ff t,0}^{-1} - \ff \Sigma)^{-1}$
where $\widetilde{\ff G}_{\ff U}[\ff \Sigma] = 
- (1/T) \delta \widetilde{F}_{\ff U}[\ff \Sigma] / \delta \ff \Sigma$.
If this was equivalent with the DMFT self-consistency condition, 
a local self-energy would be a solution.
This would imply, however, that $\widetilde{\ff G}_{\ff U}[\ff \Sigma]$ is non-local
for a local $\ff \Sigma$. 
Hence, $\widetilde{F}_{\ff U}[\ff \Sigma]$ cannot be the Legendre transform of 
$\widetilde{\Phi}_{\ff U}[\ff G]$ where $\widetilde{\Phi}_{\ff U}[\ff G]$ 
(as above) is
the Luttinger-Ward functional with vertices restricted to a single site.
An alternative choice for $\widetilde{F}_{\ff U}[\ff \Sigma]$, however, does not
suggest itself.

One can conclude that a functional of the self-energy is necessary and sufficient to 
obtain the DMFT as a type-III approximation while a functional of the Green's function 
is necessary and sufficient to get the DMFT as a type-II approximation.
The decisive point is that rather a local self-energy can be tolerated as an 
approximation than a local Green's function.

%*******************************************************************************
\section{New Approximations}
%*******************************************************************************
 
\begin{figure}[t]
\centerline{
\includegraphics[width=.6\textwidth]{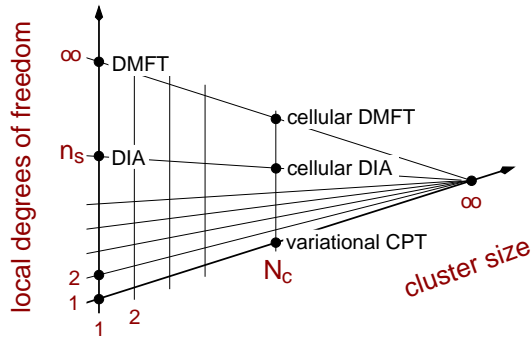}
}
\caption[]{Different possible approximations within the SFA (see text).}
\label{fig:app}
\end{figure}

The immediate return of these considerations is a number of non-perturbative and 
thermodynamically consistent type-III approximations as shown in Fig.\ \ref{fig:app}.
These differ from the DMFT by a different restriction of the domain for the self-energy
functional (\ref{eq:sfa}), i.e.\ by a different reference system with 
a different subspace of variational parameters $\ff t'$ but the same interaction 
($\ff U'=\ff U$).
The evaluation of a type-III approximation requires the repeated computation of the 
grand potential and the Green's function or self-energy of the reference system to
get $\Omega_{\ff t, \ff U}[\ff \Sigma_{\ff t',\ff U}]$ which must be optimized with
respect to $\ff t'$ subsequently.

The class of possible reference systems is essentially spanned by two parameters,
namely $n_{\rm s}-1$ which is the number of additional bath sites per correlated site
and $N_{\rm c}$ which is the number of correlated sites in a cluster that is decoupled 
from the rest of the correlated sites (Fig.\ \ref{fig:app}).
The DMFT is obtained for $N_{\rm c}=1$ and $n_{\rm s} = \infty$ since a continuous bath 
($n_{\rm s} = \infty$) is necessary to represent an arbitrary local free Green's function.

%*******************************************************************************
%*******************************************************************************

The choice $N_{\rm c}=1$ but $n_{\rm s} < \infty$ yields new approximations 
(``dynamical impurity approximations'', DIA) which are inferior as compared 
to the full DMFT but allow for much simpler and faster calculations when 
$n_{\rm s}$ is small. 
The most simple but non-trivial approximation ($n_{\rm s} = 2$-DIA) has been
shown \cite{Pot03a,Pot03b} to already cover the essence of the DMFT scenario 
\cite{GKKR96} for the Mott metal-insulator transition in the Hubbard model.
At the critical point for $T=0$ the calculations can be done even analytically
\cite{Pot03b}, and with increasing $n_{\rm s}$ the grand potential, static 
quantities and the entire phase diagram rapidly converge to the full DMFT results 
\cite{Pot03a,Poz04}.
The DIA is similar but superior as compared to the exact-diagonalization approach 
\cite{GKKR96}. Even for small $n_{\rm s}$ the approach is thermodynamically 
consistent and, off half-filling, respects the Luttinger sum rule, for example.
The DIA has also been employed successfully to study the influence of phonons 
on metal-insulator transitions in the Holstein-Hubbard model \cite{KMOH04,KMHO04}.

Nothing new is obtained for $n_{\rm s} = \infty$ and $N_{\rm c}>1$: Here the SFA
recovers the cellular DMFT \cite{KSPB01}.
(Note that the dynamical cluster approximation \cite{HTZ+98} is a type-II approximation).
More interesting is the case $n_{\rm s} = 1$ and $N_{\rm c}>1$ which turns 
out \cite{PAD03} to represent a variational generalization of the cluster-perturbation 
theory \cite{SPPL00}.
This V-CPT is well suited to describe phases with spontaneously broken symmetry and has 
been employed to study one-particle excitations and antiferromagnetic order in the 
$D=2$ and $D=1$ Hubbard model at half-filling \cite{DAH+04} and charge ordering in 
the extended Hubbard model \cite{AEvdLP+04}.
A further application concerns antiferromagnetism in quarter-filled ladder systems 
\cite{ASE04}.
An impressing example of the power of the V-CPT approach has been given recently in a 
study of the competition between antiferromagnetism and d-wave superconductivity in the 
hole- and electron-doped Hubbard model \cite{SLMT04}.
The question of phase separation is addressed in Ref.\ \cite{AA05}.

Summing up, 
the SFA is able to unify different cluster theories and local approximations within a 
single and consistent framework which offers a large flexibility in the use of bath 
sites, ficticious fields, boundary conditions and particle reservoirs \cite{PAD03}.
The formalism provides a controlled compromise between the demands for a non-perturbative 
and systematic theory working in the thermodynamic limit on the one hand, and the 
limited computational capabilities to diagonalize finite-size systems on the other.

%*******************************************************************************
\section{Open Problems}
%*******************************************************************************

The self-energy-functional approach allows to construct a series of consistent
approximations which improve systematically as $N_{\rm c} \to \infty$.
It is by no means clear, however, whether bath sites $n_{\rm s} > 1$ help to
speed up the convergence with respect to $N_{\rm c}$ and whether a cluster 
extension of DMFT or the V-CPT is more efficient.
This can be clarified only empirically by considering different lattice models in 
different dimensions.
As a few bath sites have turned out to be sufficient for reproducing the essential
mean-field ($N_{\rm c}=1$) physics in a number of studies of the single-band Hubbard
model, further applications of 
the DIA are worthwhile to explore e.g.\ the mean-field phase diagrams of more 
complex (multi-orbital) models.
Furthermore, one may also envisage the application of a simplified DMFT where a 
single (but continuous, $n_{\rm s} = \infty$) bath is optimized for a multi-orbital
model. 
This might be well justified for not too low temperatures.

On the technical side, there are two main future tasks:
The full diagonalization and the Lanczos method which have been used so far, should
be supplemented by a ``reference system solver'' based on stochastic techniques
to improve the scaling of the numerical effort with the system size.
Secondly, it would be advantageous to have an iterative technique at hand that 
directly yields a solution of the SFA Euler equation without the need for 
numerical differentiation. 
First results using full diagonalization \cite{Poz04} are encouraging.

On the conceptual side, the question for the possibility to give strict upper 
bounds to the grand potential is still open.
Probably, a positive answer requires to establish a link to the Ritz variational
principle.
On the other hand, no example is known yet where the SFA grand potential at a
stationary point is {\em lower} than the exact one.

There are different directions into which the formalism may be extended.
As the coherent-potential approximation for the disorder Anderson model has the
same (mean-field) status as the DMFT for the Hubbard model, it suggests itself that
a self-energy-functional approach can also be constructed for systems with disorder 
(and interaction). First applications \cite{BP05} demonstrate that such a theory 
\cite{PB05} is feasible.
A challenge consists in the extension of the theory to include two-particle 
Green's functions in a generalized variational principle.
Here the recently proposed functional-integral formulation of the SFA \cite{Pot04}
can be helpful.
Two-particle correlation functions are interesting by themselves and may furthermore
facilitate an even greater flexibility in the choice of reference systems.
At the same time such an approach could provide a conceptual clear way to treat
models with non-local interactions.
Currently, this problem is circumvented by a more pragmatic decoupling 
procedure \cite{AEvdLP+04}.
\newline

The author would like to thank M. Aichhorn, E. Arrigoni, F.F. Assaad, M. Balzer, 
R. Bulla, C. Dahnken, W. Hanke, A. Millis, and W. Nolting for discussions.
Support by the Deutsche Forschungsgemeinschaft within the Sonderforschungsbereich 410 
and the Forschergruppe 538 is acknowledged.

\end{document}